\documentclass[aps,prb,twocolumn,showpacs,amsmath,amssymb,amstex,mathfonts,superscriptaddress,float]{revtex4-1}

\usepackage{color,amsfonts,graphicx,verbatim}
\usepackage{amsmath,amssymb,amsfonts,amsthm}
\usepackage{amsmath}
\usepackage{bm}
\usepackage{amssymb}
\usepackage{lipsum}
\usepackage{times}
\usepackage{dcolumn}
\usepackage{cases}
\usepackage{braket}
\usepackage{booktabs}
\usepackage{setspace} 
\usepackage{}
\usepackage[caption=false]{subfig}
 
\usepackage{amsthm}
\usepackage{listings}
\usepackage{enumerate}
\usepackage{latexsym}
\usepackage{psfrag}
\usepackage{hyperref} 
 
\usepackage{array}
\usepackage{float}
\usepackage{blindtext}

\providecommand{\R}{{\mathbb{R}}}
\providecommand{\fv}{\mathbf{f}}
\providecommand{\rv}{\mathbf{r}}

\def\bal {\begin{align}}
\def\eal {\end{align}}
\def\be {\begin{equation}}
\def\ee {\end{equation}}
\def\bea {\begin{eqnarray}}
\def\eea {\end{eqnarray}}
\def\bra {\langle}
\def\ket {\rangle}

\renewcommand{\vec}[1]{\mathbf{#1}}

\newcommand{ \ep }{\varepsilon}

\newcommand{ \ff }{{f}}

\makeatletter
    \renewcommand\@make@capt@title[2]{%
     \@ifx@empty\float@link{\@firstofone}{\expandafter\href\expandafter{\float@link}}%
      {\textbf{#1}}\@caption@fignum@sep#2\quad}%
    \makeatother
   
\makeatletter 
\renewcommand{\fnum@figure}{\textbf{Figure~\thefigure}}
\makeatother

\begin{document}

\title{The adiabatic strictly-correlated-electrons functional: kernel and exact properties} 
\author{Giovanna Lani}\affiliation{Department of Theoretical Chemistry and Amsterdam Center for Multiscale Modeling, FEW, Vrije Universiteit, De Boelelaan 1083, 1081HV Amsterdam, The Netherlands}
\author{Simone di Marino}\affiliation{Laboratoire de Math\'ematiques d'Orsay, Univ. Paris-Sud, CNRS, Universit\'e Paris-Saclay, 91405 Orsay, France.} \author{Augusto Gerolin} \affiliation{Dipartimento di Matematica, Univerisit\'{a} di Pisa, Largo B. Pontecorvo, 56126 Pisa, Italy}
\author{Robert van Leeuwen} \affiliation{Department of Physics, Nanoscience Center, University of Jyv\"askyl\"a, 40014 Jyv\"askyl\"a,  Finland; European Theoretical Spectroscopy Facility (ETSF)}
\author{Paola Gori-Giorgi} \affiliation{Department of Theoretical Chemistry and Amsterdam Center for Multiscale Modeling, FEW, Vrije Universiteit, De Boelelaan 1083, 1081HV Amsterdam, The Netherlands}

\date{\today}
\begin{abstract}
We investigate a number of formal properties of the adiabatic strictly-correlated electrons (SCE) functional,  relevant for time-dependent potentials and for kernels in linear response time-dependent density functional theory. Among the former, we focus on the compliance to constraints of exact many-body theories, such as the generalised translational invariance and the zero-force theorem.
Within the latter, we derive an analytical expression for the adiabatic SCE Hartree exchange-correlation kernel in one dimensional systems, and we compute it numerically for a variety of model densities. We analyse the non-local features of this kernel, particularly the ones that are relevant in tackling problems where kernels derived from local or semi-local functionals are known to fail.
\end{abstract}
\maketitle

%
%
\section{Introduction}
While a considerable amount of work 
on the strictly-correlated-electrons (SCE) formalism \cite{Sei-PRA-99,SeiGorSav-PRA-07,GorVigSei-JCTC-09} within the framework of ground state Kohn-Sham (KS) density functional theory (DFT) has been carried out, \cite{MalMirCreReiGor-PRB-13,MenMalGor-PRB-14,MalMirGieWagGor-PCCP-14,MalMirMenBjeKarReiGor-PRL-15,VucWagMirGor-JCTP-15} the study of its performances in the time domain is just starting. \cite{Mir-THESIS-15, MirDegRubGor-XXX-16}
The aim of this work is to begin a systematic investigation of the SCE functional in the context of time dependent problems, in order to understand its fundamental aspects and its potential in tackling challenging problems for the standard approximations employed in time-dependent (TD) DFT.   

We will hence focus on those physical situations described by an explicitly time-dependent Hamiltonian, and whose dynamics is described by the time-dependent Schr\"{o}dinger equation (TDSE). Due to the existence of a time-dependent density-potential mapping \cite{RunGro-PRL-84,Leu-PRL-99,RugLeu-EPL-11,RugPenLeu-JPCM-15} for interacting and non-interacting systems, a time-dependent Kohn-Sham approach 
can be rigorously set up and employed to study the dynamics of quantum systems at a manageable computational cost.
Choosing the initial non-interacting wave function to be a single Slater determinant of some spin orbitals $\psi_j(\vec{x},t_0)$, one can reduce the TDSE to a set of single-orbital equations, the time-dependent Kohn-Sham (TDKS) equations, of the form (in Hartree atomic units used throughout):
%
%
\bea
\label{Eq:tdks}
\displaystyle
i \partial_t \psi_j(\vec{x},t) &=&  \Big{(} -\frac{1}{2} \nabla^2 + v_{ext}(\vec{r},t) + v_{\rm H}([n], \vec{r},t) \\
&+& v_{xc}([\Psi_0,\Phi_0,n];\vec{r},t) \Big{)} \psi_j(\vec{x},t), \nonumber
\eea
with $\Psi_0$ and $\Phi_0$ initial states of the true interacting and of the non-interacting KS system. The time-dependent density is thus computed in the familiar way (for simplicity in this introduction we consider closed-shell systems) as:
\be
\displaystyle
n(\vec{r},t) = 2\sum^{N/2}_{j=1} |\psi_j(\vec{r},t)|^2.
\label{Eq:td-dens}
\ee
In Eq.~\eqref{Eq:tdks}, $v_{\rm H}([n], \vec{r},t)$ is the usual Hartree potential computed with the time-dependent density $n(\vec{r},t)$, and $v_{xc}([\Psi_0,\Phi_0,n];\vec{r},t) $ is the exchange-correlation (xc) potential, depending also on the initial states $\Psi_0$ and $\Phi_0$.
Trading the many-body TDSE for the one-particle TDKS equations has a price to pay, that is the time-dependent exchange-correlation potential $v_{xc}([n, \Psi_0,\Phi_0];\vec{r},t)$ of TDDFT is an even more complex object than the $v_{xc}$ for ground state DFT, as it is a functional of the density at all times $t'\le t$ and, additionally, of the initial state of both the interacting and non-interacting systems. 
However, whenever the initial state for the evolution problem described by the TDKS equations is chosen to be the ground state of the system, then the functional dependence of the xc potential is on the electronic density alone: since this scenario occurs naturally in many problems of interest, it doesn't pose actual limitations and thus it is often adopted in practical applications.

Similarly to ground state DFT, in order to make use of Eq.~(\ref{Eq:tdks}) one needs approximations for the exchange correlation potential \( v_{xc} \). 
%
A first drastic approximation, which is used in the large majority of cases in TDDFT, is the so-called \textit{adiabatic approximation}, obtained by inserting in a ground-state approximate \( v_{xc}([n];\vec{r},t) \) the \textit{instantaneous} density, ignoring dependence on the density at earlier times.
This approximation has a very specific range of validity -- infinitely slowly varying perturbations, such that the system is always in its ground state -- but it is very often employed outside it, with results that can vary from very satisfactory to poor, depending on the nature of the problem addressed.
In certain cases, it is still difficult to disentangle the errors due to the adiabatic approximation and the errors due to the approximation for the ground-state exchange correlation potential, but considerable progress has been made in recent years, by analysing, when possible, the ``adiabatically exact'' potential. \cite{FukFarTokAppKurRub-PRA-13,EllFukRubMai-PRL-12,FukMai-PCCP-14}
In other cases, it is instead well established that neglecting all ``memory effects'' in \( v_{xc} \) (or equivalently frequency dependence in the so called xc kernel, \( \mathcal{F}_{xc} \) of linear response TDDFT), 
does not allow TDDFT to describe excitations with a predominantly double character. \cite{HirHea-CPL-99,ThiKue-PRL-14,TozHan-PCCP-00,NeuBaeNoo-JCP-04, MaiZhaCavBur-JCP-04} In the TDDFT framework, adiabaticity is thus equivalent to locality in time.
The most common approximations used to build the adiabatic  \( v_{xc} \) (and \( \mathcal{F}_{xc} \) in the linear response case) in TDDFT, are local and semi-local functionals, which thus add to locality in time locality in space as well. In these cases, the time-dependent kernel \( \mathcal{F}_{xc}(\rv,\rv',t,t') \) is approximated as
\begin{equation}
	\mathcal{F}_{xc}([n];\rv t,\rv't')=\frac{\delta^2 E_{xc}^{\rm
            appr}[n]}{\delta n(\rv,t) \delta n(\rv',t')} ,
\end{equation}
where $E_{xc}^{\rm appr}[n]$ is evaluated at the instantaneous density
and it is often a local or semi-local approximate
functional, which makes the kernel different from zero only on (or very close to) the diagonal $\rv=\rv'$.
 We have already hinted at the shortcomings of the locality in time in this introduction, but also the locality in space has serious limitations, a notorious example being the description of excitations with a long-range charge transfer (CT) character \cite{DreHea-JACS-04, BaeGriMee-PCCP-13}. In the case of closed-shell fragments, the introduction of a considerable portion of Hartree-Fock exchange (often introduced at long-range only through range-separation) is able to fix the CT problem in linear response TDDFT. \cite{BanGri-CTC-14, DieGri-JPCA-04, PeaBenHelToz-JCP-12, YanTewHan-CPL-04,DreWeiHea-JCP-03} However, this solution does not work for the very challenging case of homolytic bond breaking excitations, the prototypical example being the lowest excited singlet state $^1\Sigma_u^+$ of the H$_2$ molecule. \cite{GriGisGorBae-JCP-00,GieBae-CPL-08} In this case, the kernel should diverge in order to compensate the fact that this excitation in the KS system goes to zero as the bond is broken. \cite{GriGisGorBae-JCP-00,GieBae-CPL-08} In this context, we will show that, at least in a model one-dimensional case, the adiabatic SCE (ASCE) kernel shows a very promising non-local diverging behavior. \\
In order to construct approximations both for potentials and kernels in TDDFT, one can be guided by trying to satisfy exact properties and constraints of many-body theories. In a series of works \cite{Dob-PRL-94,Vig-PRL-95,Vig-PLA-95} Dobson and Vignale devised a number of constraints (named theorems afterwards) that the time-dependent $v_{xc}$ should comply to, in order to avoid unphysical results or contradictions in the theory. 
From their analysis, it appeared for the first time that the interplay between non locality in space and non locality in time is a delicate issue in TDDFT and this fact needs to be kept in mind when looking for approximations, making this task much more challenging than in ground state DFT. It is thus natural to ask whether a highly non-local functional such as SCE can satisfy these exact conditions when employed in the adiabatic approximation.

After briefly reviewing in Sec.~\ref{sec:rev} the basics ideas of the SCE formalism, we will show in Sec.~\ref{sec:exact-prop} how the SCE potential satisfies exact properties of many-body theories, such as the zero-force theorem and the generalized translational invariance. Our analysis will also show how, while non-locality in time and non-locality in space have to go hand in hand, non-locality in space and locality in time can coexist without violating the above mentioned properties. 
In Sec.~\ref{sec:kernel}  we will derive an analytical expression for the SCE kernel for one-dimensional systems, and then compute it numerically for various density profiles. We will complete the section with a discussion on some general features of the kernel, pinpointing at those which arise from its highly non-local nature and which could be promising for the description of bond-breaking excitations.
%
Finally we will give our conclusions and perspectives for future work.
%
%
\section{Review of the SCE formalism}
\label{sec:rev}
The SCE formalism can be put in the DFT context starting with the generalization of the Hohenberg-Kohn functional $F[n]$ to scaled interactions:
\begin{equation}
\displaystyle
F_{\lambda}[n] = \min_{\Psi \rightarrow n } \bra \Psi | \hat{T} + \lambda \hat{V}_{ee} | \Psi \ket
\label{Eq:ghk-func}
\end{equation}
where $\hat T$ and $\hat V_{ee}$ are the familiar kinetic and two-body interaction operators, while $\lambda$ is a parameter varying continuously from 0 to $\infty$, yielding different scenarios: $F_{\lambda=0}[n]=T_s[n]$ corresponds to the non interacting or Kohn-Sham system, $F_{\lambda=1}[n]$ corresponds to the real physical system,  while $ F_{\lambda=\infty}[n]$ defines the strong-coupling limit, \cite{Sei-PRA-99,SeiGorSav-PRA-07} captured by the strictly-correlated-electron functional
\begin{equation}
 \displaystyle
 V^{\rm SCE}_{ee}[n] \equiv \min_{\Psi \rightarrow n} \bra \Psi |\hat{V}_{ee} | \Psi \ket.
 \label{Eq:min-sce}
\end{equation}
The working hypothesis to build the minimizer of Eq.~\eqref{Eq:min-sce} for a given density is that in this limit the many-body wavefunction collapses into a 3-dimensional subspace of the full configuration space,
\begin{multline}
\label{eq_psi2}
|\Psi_{\rm SCE}(\rv_1,\dots,\rv_N)|^2 = \frac{1}{N!} \sum_{\wp}
\int d\rv \, \frac{n(\rv)}{N} \, \delta(\rv_1-\fv_{\wp(1)}(\rv)) \\
\times\delta(\rv_2-\fv_{\wp(2)}(\rv)) \cdots \delta(\rv_N-\fv_{\wp(N)}(\rv))\; ,
\end{multline}
where $\wp$ denotes a permutation of ${1,\dots,N}$, such that 
$n(\rv) = N \int |\Psi_{\rm SCE}(\rv,\rv_2,\dots,\rv_N)|^2 
\,d\rv_2\cdots d\rv_N$. The functional is then specified in terms of the so-called co-motion functions $\vec{f}_i([n]; \vec{r})$ that determine the set in which $|\Psi_{\rm SCE}|^2\neq 0$,
\begin{equation}
V_{ee}^{\rm SCE}[n]=\frac{1}{2}\int n(\rv)\sum_{i=2}^N\frac{1}{|\rv-\vec{f}_i([n]; \vec{r})|}d\rv.	
\end{equation}
The functional derivative of $V_{ee}^{\rm SCE}[n]$ defines the SCE potential:
\begin{equation} 
 \displaystyle
 v^{\rm SCE}([n];\vec{r}) = \frac{\delta V^{\rm SCE}_{ee}[n]}{\delta n(\vec{r})}, 
 \label{Eq:sce_def}
\end{equation}
which can be computed via a rigorous and physically transparent shortcut \cite{SeiGorSav-PRA-07} as the repulsion felt by an electron in $\rv$ due to the other $N-1$ electrons at positions $\rv_i=\vec{f}_i([n]; \vec{r})$,
\begin{equation}
\displaystyle
\nabla v^{\rm SCE}([n]; \vec{r}) = -\sum^{N}_{i=2} \frac{\vec{r} - \vec{f}_i([n]; \vec{r})}{|\vec{r} - \vec{f}_i([n]; \vec{r})|^3}.
\label{Eq:3dsce-pot}
\end{equation}
%
%
All the $\vec{f}_i[n](\vec{r})$, whose physical meaning is to give the positions of all the others $N-1$ electrons once the position of a reference electron has been fixed in $\vec{r}$, satisfy the following non-linear differential equation:
\begin{equation}
\displaystyle
n(\vec{f}_i([n];\vec{r})) d \vec{f}_i([n];\vec{r}) = n(\vec{r}) d\vec{r} \quad \quad i = 2,..,N-1
\label{Eq:diffeqf}
\end{equation}
which shows their non-local dependence on n$(\vec{r})$.
Furthermore the co-motion function obey (cyclic) group properties \cite{SeiGorSav-PRA-07} which ensure that the electrons are indistinguishable.

In the recent years, it has been realized that the problem defined by the minimization \eqref{Eq:min-sce} is equivalent to an optimal transport problem with Coulomb cost. \cite{ButDepGor-PRA-12,CotFriKlu-CPAM-13}
Since then, the optimal transport community has been able to prove several rigorous results. In particular, the SCE state \eqref{eq_psi2} has been proven to be the true minimizer for any number of particles $N$ in one dimensional (1D) systems \cite{ColDepDiM-CJM-14} and in any dimension for $N=2$. \cite{ButDepGor-PRA-12} For more general cases, it has been shown that the minimizer might not be always of the SCE form \cite{ColStr-arxiv-15}. Even in those cases, however, SCE-like solutions seem to be able to go very close to the true minimum, \cite{DiMNenGor-XXX-15} and in several cases it is still possible to prove Eq.~\eqref{Eq:3dsce-pot} . \cite{DiMNenGor-XXX-15}

In the low-density limit (or strong-coupling limit) the exact Hartree and exchange-correlation (Hxc) energy functional of KS DFT tends asymptotically to $V_{ee}^{\rm SCE}[n]$, \cite{MalGor-PRL-12,CotFriKlu-CPAM-13} Thus, in the following we denote $v^{\rm SCE}([n]; \vec{r})$ of Eq.~\eqref{Eq:3dsce-pot} as $v_{\rm Hxc}^{\rm SCE}([n]; \vec{r})$, to stress that this potential is the strong-coupling approximation to the standard Hxc potential of KS DFT. \cite{MalGor-PRL-12,MalMirCreReiGor-PRB-13}
%
%
%
%
%
%
%
%

Now that the basics of the SCE formalism at the ground state level have been reviewed, we can move to the time-dependent domain. 
%
%
\section{Exact properties from many-body theories}
\label{sec:exact-prop}
%
%
Since the success of TDDFT relies heavily on the availability and the quality of the approximations for $v_{xc}([n];\vec{r},t)$ and for the linear response exchange-correlation kernel $\mathcal{F}_{xc}([n],\vec{r},\vec{r}',t,t')$, there have been intense research efforts towards better approximations.
\noindent As already mentioned in the introduction, a way to guide such approximations is to resort to the compliance to exact constraints from many-body theories, similarly to what has been done extensively already in ground state DFT. 
%
A first exact condition is given by scaling relations, \cite{HesParBur-PRL-99, HesMaiBur-JCP-02} a second one by a sum rule for the time-dependent exchange-correlation energy \cite{HesParBur-PRL-99} and just like in the static case, the time-dependent xc potential should be self-interaction free.\\
In addition to the constraints enumerated above, a very important condition on approximate xc potentials is that they should be \textit{Galilean invariant}, as a consequence of the fact that the TDSE itself exhibits this symmetry.  
This condition was first investigated by Vignale, \cite{Vig-PRL-95} as a generalization of an earlier work by Dobson \cite{Dob-PRL-94} on the so called harmonic potential theorem (HPT), which states that upon the application of a time-dependent field to a many-body system confined by an harmonic potential, its time-dependent density is rigidly shifted.
In \cite{Vig-PRL-95} it was demonstrated that the HPT is automatically satisfied whenever the time-dependent xc potential obeys a precise constraint, that is upon a rigid shift of the system's time-dependent density, the time-dependent xc potential is rigidly translated by the same quantity. We will refer to this property as generalized translational invariance (GTI), since it holds also for coordinates frames which are accelerated with respect to the original one. \footnote{For frames which are translated with a uniform velocity with respect to the reference one, we simply talk about Galilean invariance} \\
%
%
%
%
%
Thus in general the GTI can be formalized as follows: given an arbitrary (be or not time-dependent) 
shift of the density $\vec{R}(t)$:
\begin{equation}
\displaystyle
  n'(\vec{r},t) = n(\vec{r} - \vec{R}(t),t)
\label{Eq:denshift}
\end{equation}
the xc potential associated with this density has to transform accordingly to:
\begin{equation}
\displaystyle
v_{xc}([n'];\vec{r},t) = v_{xc}([n];\vec{r} -\vec{R}(t),t)  
\label{Eq:gtigen}
\end{equation}
%
%
%
%
%
\subsection{Properties of the adiabatic SCE potential}
We will now show explicitly show that the ASCE complies to this requirement. \\
We begin by observing that in the SCE limit, upon the shift of the density, all the relative distances between the electrons have still to be the same, thus the co-motion functions transform as:
%
%
\begin{equation}
\displaystyle
\vec{f}_i([n']; \vec{r}) = \vec{f}_i([n]; \vec{r} - \vec{R}(t)) + \vec{R}(t).
\label{Eq:transf}
\end{equation}
In the one dimensional case, where the co-motion functions can be expressed in terms of a simple one-dimensional integral, one can show explicitly that the above relation holds, see App.~\ref{App:1} for details.
%
Substituting the transformed co-motion functions into the expression for the SCE potential gives:
\begin{eqnarray}
\nabla v^{\rm SCE}_{\rm Hxc}([n'];\vec{r},t) &=& 
%
  -\sum^{N}_{i=2} \frac{\vec{r} - \vec{f}_i([n]; \vec{r} - \vec{R}(t)) - \vec{R}(t)}{|\vec{r} - \vec{f}_i([n]; \vec{r} - \vec{R}(t)) - \vec{R}(t)|^3} \nonumber \\
&=& \nabla v^{\rm SCE}_{\rm Hxc}([n];\vec{r} -\vec{R}(t),t). 
\end{eqnarray}
%
Integration and subtraction of the Hartree potential (which satisfy the GTI straightforwardly) yields:
\begin{equation}
\displaystyle
v^{\rm SCE}_{xc}([n'];\vec{r},t) =  v^{\rm SCE}_{xc}([n];\vec{r} -\vec{R}(t),t)
\label{eq:SCEpotGTI}
\end{equation}
which is the relation we wanted to prove. \\
A second important constraint is that the xc potential cannot exert a net external force on the system, which is nothing else than the compliance to Newton's third law of motion. In DFT this property goes under the name of zero force theorem (ZFT) and in \cite{Vig-PRL-95} it was shown how it is automatically satisfied for translationally invariant xc potentials.
One may think that this is a trivial requirement to be satisfied, but in practice it isn't.
For example in \cite{MunKumLeeuRei-PRA-07} it was demonstrated numerically that computing the dipole moment of small Na$_5$ and Na$_9^+$ clusters, via the exact exchange Krieger-Li-Iafrate approximation to $v_{xc}$, yielded an increased amplitude in the dipole oscillations, most likely due to spurious internal forces appearing as a consequence of the violation of the ZFT.
%
The ZFT not only has implications for the approximations to the time-dependent xc potential, but also on another key quantity of TDDFT, namely the exchange correlation kernel. In \cite{Vig-PLA-95} Vignale showed how a frequency dependent (thus non local in time) $\mathcal{F}_{xc}$ \textit{cannot} be local in space, in order to satisfy the ZFT. A notable example of a kernel which violates the ZFT and the HPT too, is the Gross-Kohn $\mathcal{F}_{xc}$ \cite{GroKoh-PRL-85} which indeed is frequency dependent, but local in space, as it is based on the homogeneous electron gas.
This peculiar issue in TDDFT is commonly known as \textit{ultra non-locality} problem and makes particularly challenging the construction of approximate frequency dependent kernels.
%
Adiabatic $\mathcal{F}_{xc}$, derived from fully local functionals, do not violate the ZFT. It is legitimate to ask if an adiabatic \textit{but} highly non local functional like the ASCE, does violate the ZFT. Strictly speaking we already know that it doesn't, since it respects the GTI, but in the following we will explicitly show that while non locality in time requires non locality in space, \textit{the converse is not true}. \\
%
%
Let's consider once again a shift in the density: $n'(\vec{r}) = n(\vec{r} - \vec{R})$. 
Observing that the generalized HK energy functional $F_{\lambda}[n]$ is translationally invariant (since both $\hat T$ and $\hat V_{ee}$ are) one has:
\begin{equation}
 \displaystyle
 F_{\lambda}[n] = F_{\lambda}[n'].
 \label{Eq:HKZFT}
\end{equation}
Expansion of the density in powers of $\vec{R}$ gives:
\begin{equation}
	\label{eq:expadens}
\displaystyle
n'(\vec{r}) = n(\vec{r} - \vec{R}) = n(\vec{r}) - \vec{R}\cdot \nabla n(\vec{r}) + \textbf{O}(\vec{R}^2), 
\end{equation}
and expanding both sides of Eq.~\eqref{Eq:HKZFT} yields:
\begin{eqnarray}
 \displaystyle
 0 &=& 
  \int d\vec{r} \frac{\delta F_{\lambda}}{\delta n(\vec{r})} (-\vec{R}  \cdot \nabla n(\vec{r})) 
\end{eqnarray}
%
which is valid for any arbitrary shift $\vec{R}$.\\
%
The case $F_{\lambda=\infty} = V^{\rm SCE}_{ee}$ corresponds to the SCE functional, hence:
\begin{eqnarray}
 \displaystyle
 0 &=& \int d\vec{r} v^{\rm SCE}([n]; \vec{r}) \nabla n(\vec{r}) \nonumber \\
 &=& -\int d\vec{r} \nabla v^{\rm SCE}([n]; \vec{r})\, n(\vec{r}) 
 \end{eqnarray}
which shows that the SCE potential does indeed satisfy the ZFT for \textit{static densities}. 
Additionally, since the differentiation above is completely general and holds for \textit{any} density, even time-dependent ones, one has:
\begin{equation}
 \displaystyle
 0 = \int d\vec{r} \nabla v^{\rm SCE}([n]; \vec{r},t) n(\vec{r},t), 
 \label{Eq:ZFTTD}
\end{equation}
which shows that the ASCE xc potential satisfies the ZFT for \textit{time-dependent densities} as well.\\
\subsection{Properties of the adiabatic SCE kernel}
Let's now turn to the ASCE kernel, 
\begin{equation}
	\mathcal{F}_{\rm Hxc}^{\rm ASCE}([n];\rv t,\rv't')=\frac{\delta^2 V_{ee}^{\rm SCE}[n]}{\delta n(\rv,t)n(\rv',t')}\delta(t-t'),
\end{equation}
Once again we resort to an expansion for the density in $\vec{R}(t)$, that is
$ \displaystyle n(\vec{r} - \vec{R}(t),t) \approx n(\vec{r},t) - \vec{R}(t)\cdot\nabla n(\vec{r},t) $,
%
combining this with Eq.~\eqref{eq:SCEpotGTI} and invoking the arbitrariness of $\vec{R}(t)$ and the definition of ASCE xc kernel, we obtain:
%
%
\begin{equation}
 \displaystyle
 \int d\vec{r}' \mathcal{F}_{\rm Hxc}^{\rm ASCE}([n];\vec{r}t,\vec{r}'t') \nabla n(\vec{r}') = \delta(t-t') \nabla v_{\rm Hxc}^{\rm SCE}([n]; \vec{r},t), \nonumber
\end{equation}
which shows that the ASCE kernel indeed satisfies the ZFT in the \textit{linear response} regime.
%
\subsection{Properties of the co-motion functions}
At this point it seems natural to also investigate some properties of the co-motion functions. Combining again the expansion for the density of Eq.~\eqref{eq:expadens} with Eq.~\eqref{Eq:transf} one obtains:
%
%
\begin{equation}
\int d\vec{r'} \frac{\delta \vec{f}_{i, \alpha}([n];\vec{r}) }{\delta n(\vec{r'}) } 
\, \frac{\partial}{\partial r'_{\beta}} n(\vec{r'}) =\frac{\partial}{\partial r_{\beta}}  \vec{f}_{i,\alpha}([n];\vec{r}) - \delta_{\alpha \beta}
\label{Eq:fsumrules}
\end{equation}
where $\alpha,\beta$ run over Cartesian indices $x,y,z$.
Eq.~\eqref{Eq:fsumrules} is a \textit{sum rule} that can be written also for adiabatic time-dependent co-motion functions and the static density and may be employed as constraint to devise \textit{approximate} co-motion functions.
Furthermore it can be used to verify (particularly in the easier one-dimensional case) the functional variation of the co-motion functions with respect to the density.
%
\section{SCE Hartree-exchange correlation kernel for one-dimensional systems}
\label{sec:kernel}
In the one-dimensional case with convex repulsive interparticle interaction $w(|x|)$, the SCE solution \cite{Sei-PRA-99} is known to be exact for any number of electrons $N$, \cite{ColDepDiM-CJM-14} and can be expressed in a rather simple form in terms of the function $N_e([n];x)$,
\begin{equation}
N_e([n];x)=\int_{-\infty}^xn(y) dy,
\end{equation}
and of its inverse $N_e^{-1}([n];x)$:
\be
\displaystyle
f_i([n];x) = f^{+}_i([n];x) \, \theta(x - a_k[n]) + f^{-}_i([n];x) \, \theta(a_k[n] - x) \nonumber
\label{Eq:como-1d}
\ee
where $\theta(x)$ is the usual Heaviside step function, and
\bea
\displaystyle
f^{+}_i([n];x) &=& N_{e}^{-1} \left([n]; N_{e}([n];x) +i-1 \right) \\
f^{-}_i([n];x) &=& N_{e}^{-1}\left( [n];N_{e}([n];x) +i-1 -N \right), \nonumber 
\label{Eq:como-1d-2}
\eea
with \( \displaystyle a_k[n] = N_{e}^{-1}([n];k) \), and $k = N+1-i$.

In this case the SCE potential is simply given by
\begin{multline}
v^{\rm SCE}_{\rm Hxc}([n];x) =  -\sum^{N}_{i=2} \int_x^\infty w'(|y - f_i([n];y)|)\times
\\  {\rm sgn}(y-f_{i}([n];y)) dy.
\label{Eq:1dsce-pot}
\end{multline}
The SCE kernel is then equal to the variation of the SCE potential with respect to the electron density, 
\be
\displaystyle
\mathcal{F}^{\rm SCE}_{\rm Hxc}([n];x,x')= \frac{\delta v^{\rm SCE}_{\rm Hxc}([n];x)}{\delta n(x')},
\label{Eq:1dker-def}
\ee
which can be carried out (the details of the derivation are given in Appendix~\ref{App:2}-\ref{App:varSCEpot}), yielding (for densities supported on the whole real line)
%
%
the compact expression
\bea
\displaystyle
\label{Eq:1dker-general}
\mathcal{F}^{\rm SCE}_{\rm Hxc}([n];x,x') &=& \sum^{N}_{i=2} \int^{\infty}_x \frac{w''(|y - f_i([n];y)|)]}{n(f_i([n];y))}  \\
&\times& \left[\theta(y - x') - \theta( f_i([n];y) - x')  \right] dy. \nonumber
\eea
From this expression, it is not evident that Eq.~\eqref{Eq:1dker-general} satisfies the symmetry requirement $\mathcal{F}^{\rm SCE}_{\rm Hxc}([n];x,x') =\mathcal{F}^{\rm SCE}_{\rm Hxc}([n];x',x)$. An explicit proof of this symmetry is given in Appendix~\ref{app_symmetry}.

\subsection{Analytical Example}
We begin by considering $N=2$ electrons in the Lorentzian density profile,
\begin{equation}
	\label{eq:lorentzian}
	n(x)=\frac{2}{\pi}\frac{1}{1+x^2},
\end{equation}
for which the co-motion function is simply $f_2(x)\equiv f(x)=-\frac{1}{x}$.
From the general expression of Eq.~\eqref{Eq:1dker-general},  we obtain in the first quadrant
\begin{equation}
	\label{eq:kernIandIII}
	\mathcal{F}^{\rm SCE}_{\rm Hxc}([n];x,x')=G(-\max\{x,x'\})\qquad {\rm for}\;\;x>0,\;x'>0,
\end{equation}
where we have defined the function $G(x)$
\begin{equation}
G(x)=\int_{-\infty}^x	\frac{w''(|y - f([n];y)|)]}{n(f([n];y))} dy,
\label{eq:defG}
\end{equation}
which in this case, and with e-e interaction $w(|x|)=\frac{1}{|x|}$ (since in the SCE wavefunction the particles never get on top of each other, the $1/x$ divergence at $x=0$ in 1D does not pose any problem), is equal to
\begin{equation}
	G(x)=\Bigg\{
\begin{array}{l}
\frac{\pi}{2}\frac{1}{1+x^2} \qquad  \; x\leq 0 \\
 \frac{\pi}{2}\frac{1+2 x^2}{1+x^2} \qquad x> 0.
\end{array} 
\end{equation}
Since our density satisfies $n(-x)=n(x)$, in this case the kernel in the third quadrant ($x<0$ and $x'<0$) is equal to the one in the first quadrant. In the second quadrant -- and by symmetry the fourth, since $\mathcal{F}^{\rm SCE}_{\rm Hxc}([n];x,x')=\mathcal{F}^{\rm SCE}_{\rm Hxc}([n];x',x)$ --  the kernel is given by
\begin{multline}
	\mathcal{F}^{\rm SCE}_{\rm Hxc}([n];x,x')=\left(G(x')-G(x)+G(0)\right)\theta(x'-f(x))\\ {\rm for}\;\;x>0,\;x'<0.
		\label{eq:kernIIandIV}
\end{multline}
The resulting SCE $\mathcal{F}^{\rm SCE}_{\rm Hxc}([n];x,x')$ for this case is plotted in the first panel of Fig.\ref{Fig:fig1}: as it is evident from Eq.~\eqref{eq:kernIandIII}, the kernel has in the first and third quadrants ($x,x'>0$ and $x,x'<0$) the same value as along the diagonal ($x=x'$), while in the second and fourth quadrants (Eq.~\eqref{eq:kernIIandIV}) the kernel is different from zero only in the region delimited by the $x,x'$ axes and the co-motion function $x'=f(x)$. The behavior of an adiabatic kernel local in space, such as the ALDA, is instead radically different: the $\mathcal{F}_{xc}^{\rm ALDA}(n];x,x')$ has a non-zero component only along the diagonal, $\delta(x-x')$, and the Hartree component, equal to $w(|x-x'|)$, has a maximum on the diagonal, decaying as $1/|x-x'|$ outside it. 

\subsection{Model Homonuclear molecule}
\label{Sec:num-calc.}
\begin{figure} 
 
 \subfloat[SCE kernel for a Lorentzian density.]{
  \includegraphics[clip,width=0.55\columnwidth]{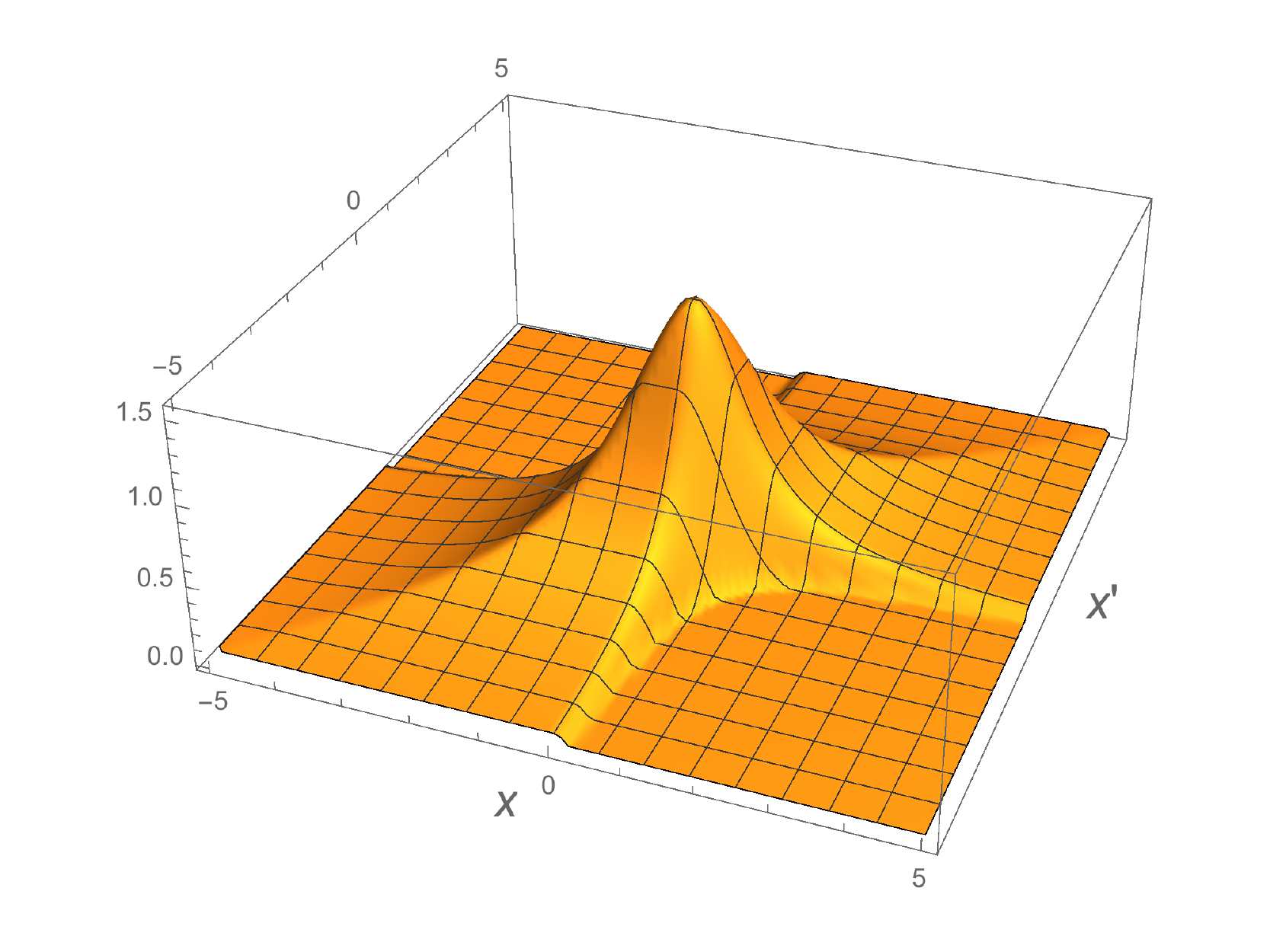}
  }
 
 \subfloat[SCE kernel for a ``homonuclear dimer'' density with \( R=3 \)]{
  \includegraphics[clip,width=0.55\columnwidth]{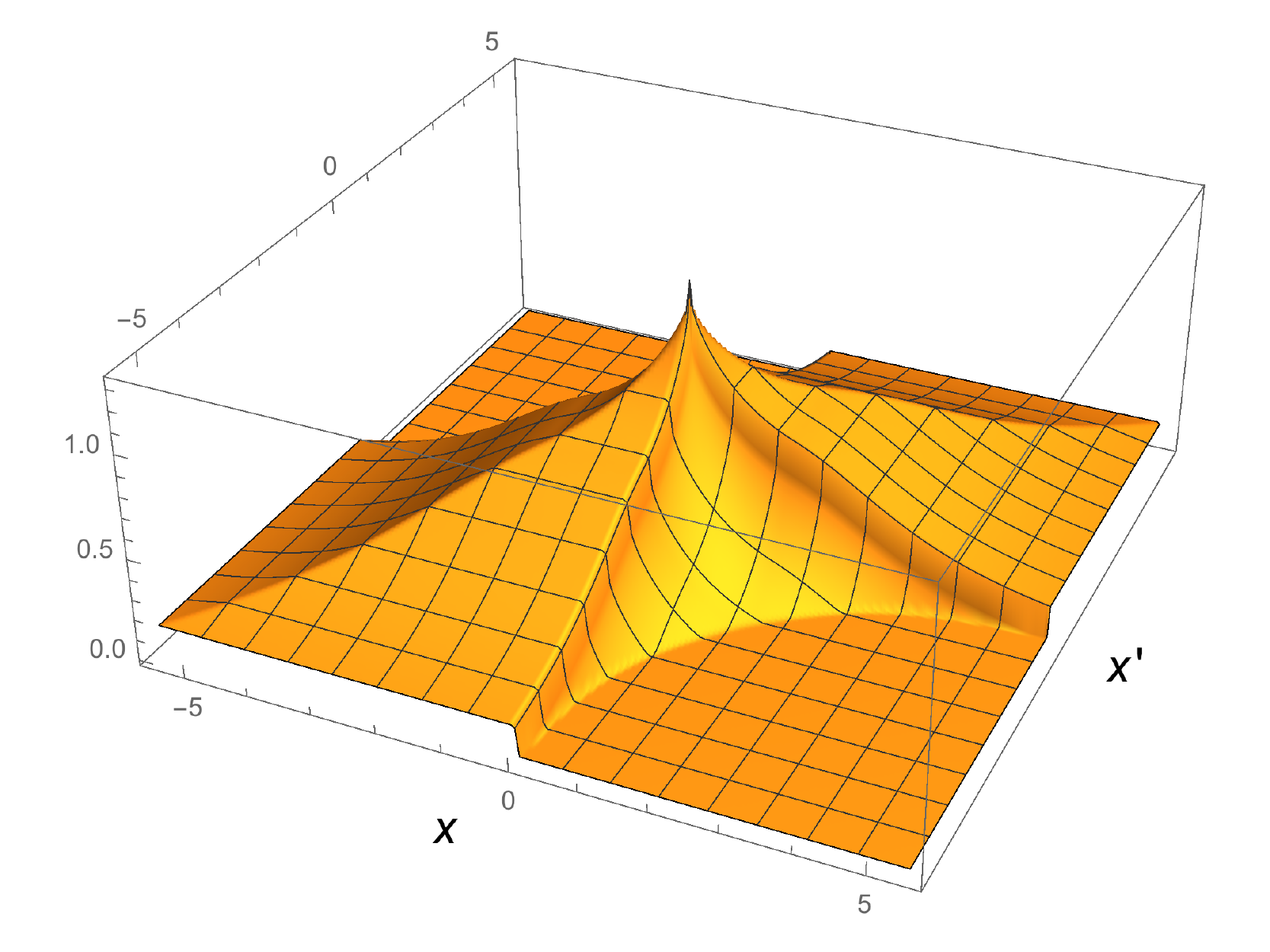}
  }
  
 \subfloat[SCE kernel for a ``homonuclear dimer'' density with \(R=8 \)]{
  \includegraphics[clip,width=0.55\columnwidth]{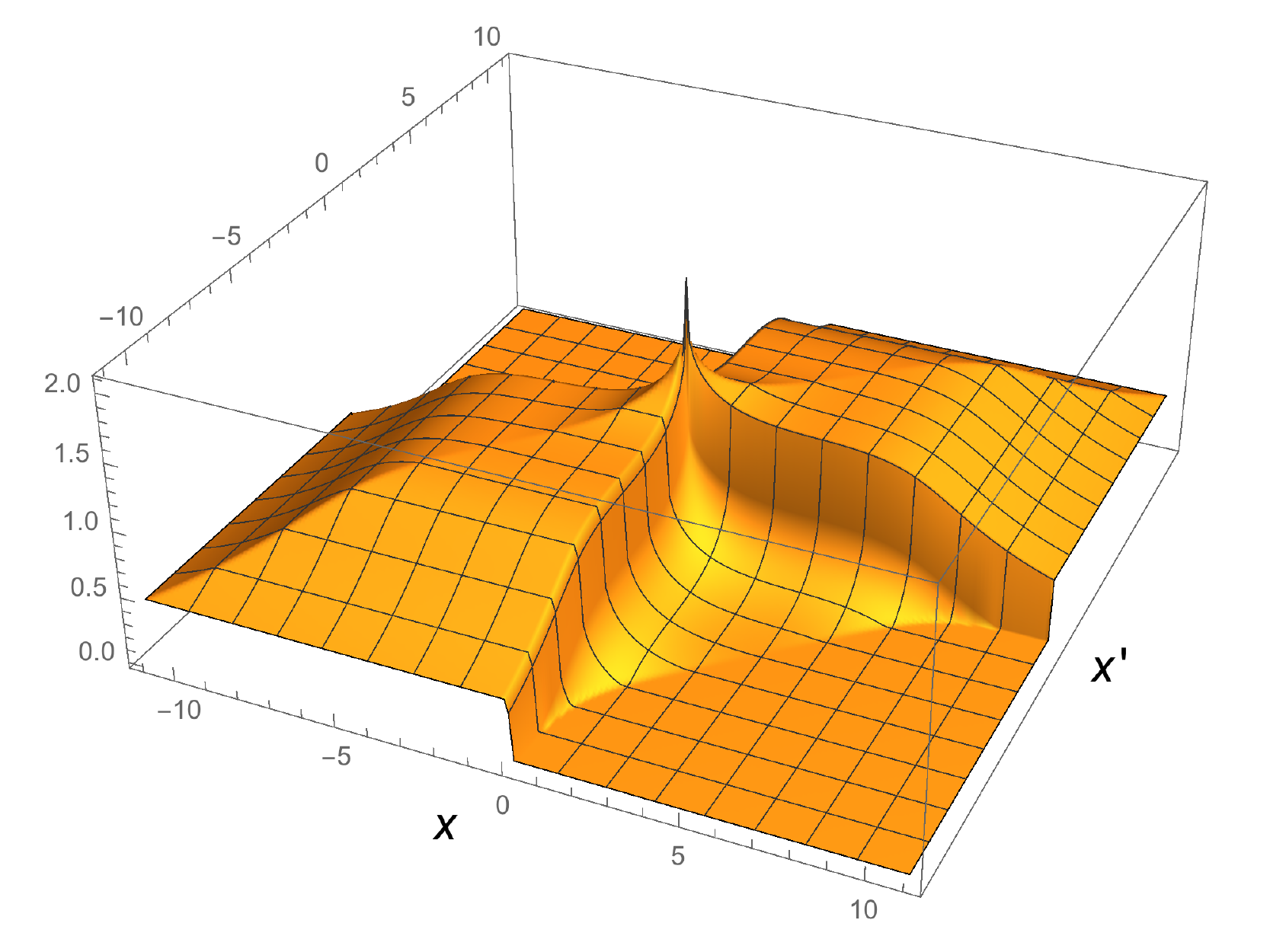}
  } 
 
 \subfloat[SCE kernel for a ``homonuclear dimer'' density with \( R=12 \)]{
  \includegraphics[clip,width=0.55\columnwidth]{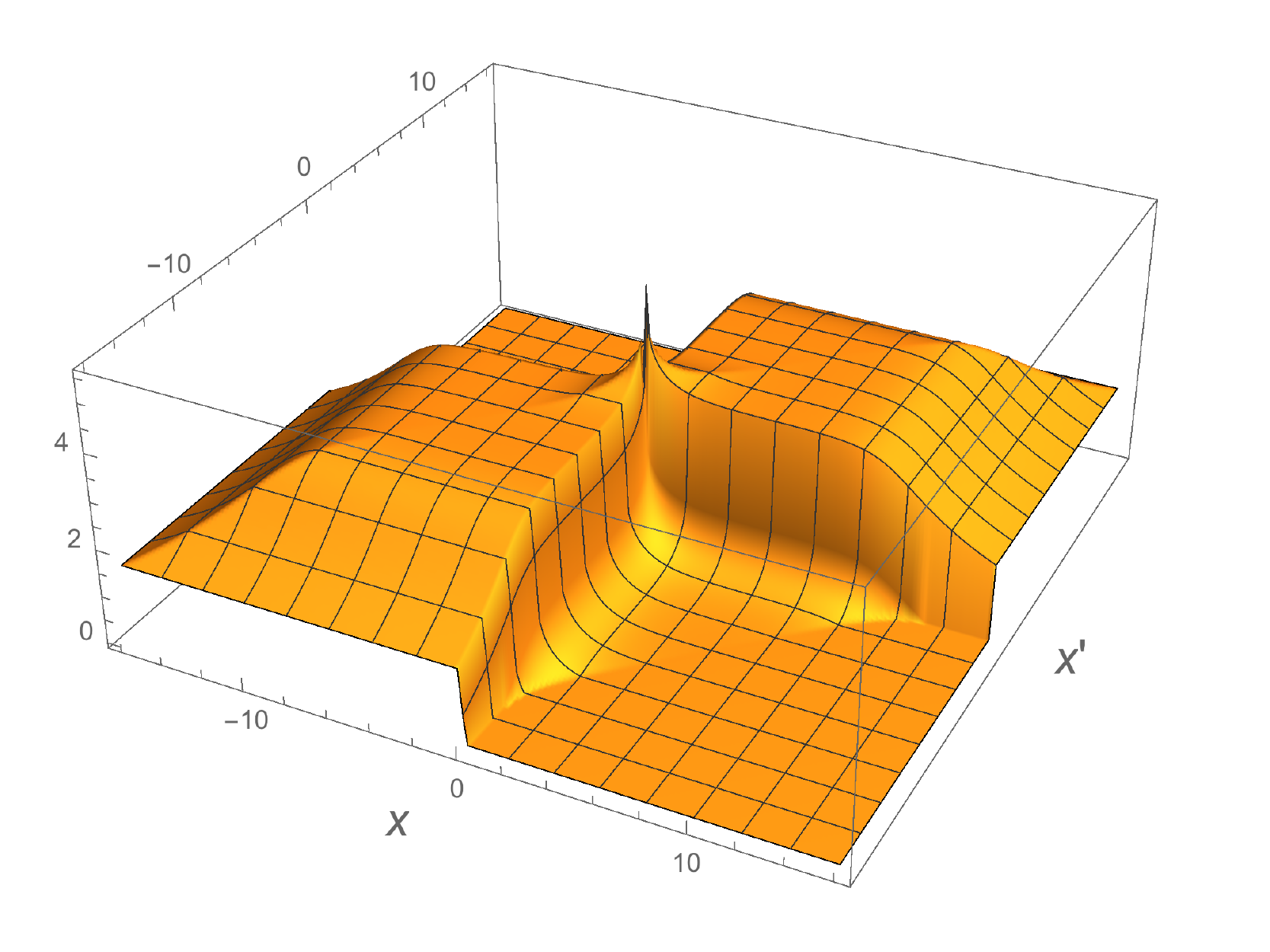}
  }
  
 \caption{The SCE Hartree-exchange-correlation kernel for $N=2$ electrons for the Lorentzian density profile of Eq.~\eqref{eq:lorentzian} (a) and for ``homonuclear dimer'' densities $n(x)=\frac{1}{2}\left(e^{-|x+R/2|}+e^{-|x-R/2|}\right)$, with \(R=3\) (b), \(R=8\) (c) and \( R=12 \) (d). The $\mathcal{F}^{\rm SCE}_{\rm Hxc}([n];x,x')$ displays, in all the four cases, a peak structure in correspondence of the vertical asymptotes of the co-motion functions. Two symmetric quasi-plateaux, in the first and third quadrants, appear for the stretched ``homonuclear dimer'' densities: they become greater in height and flatter as the value of $R$ is increased. }
\label{Fig:fig1}
 \end{figure}

The second type of density considered is a model 2-electron density which resembles the one of a homonuclear molecule,  \begin{equation}  
n(x) = \frac{a}{2} \left( e^{-a |x-\frac{R}{2}| } + e^{-a |x+\frac{R}{2}| }  \right),
\label{eq:defdensH2}
\end{equation} 
where \(a=1\) and where \( R \), the distance between the two nuclei, can be increased arbitrarily to simulate the molecular bond stretching. 
For this case we numerically computed the SCE kernel for different values of \( R \), to obtain insights on how a highly non-local kernel behaves for a problem which bears a resemblance to the H$_2$ dissociation. 
The results for \(R=3\), \(R=8\) and \(R=12\) are presented respectively in panels (b), (c) and (d) of Fig.\ref{Fig:fig1}.
%
%
%

Aside from the peak in the origin, a very interesting feature displayed by the SCE kernel is the appearance, as $R$ is increased, of two \textit{plateaux}, each occupying a large square region (of size $\approx R\times R$) of the first and the third quadrants.
As in the case of the lorentzian density, the SCE kernel has also non-zero components in the second and fourth quadrants, but they are now much smaller than the ones in the I and III quadrants.

\begin{figure}[!htbp]
 \includegraphics[width=0.9\columnwidth]{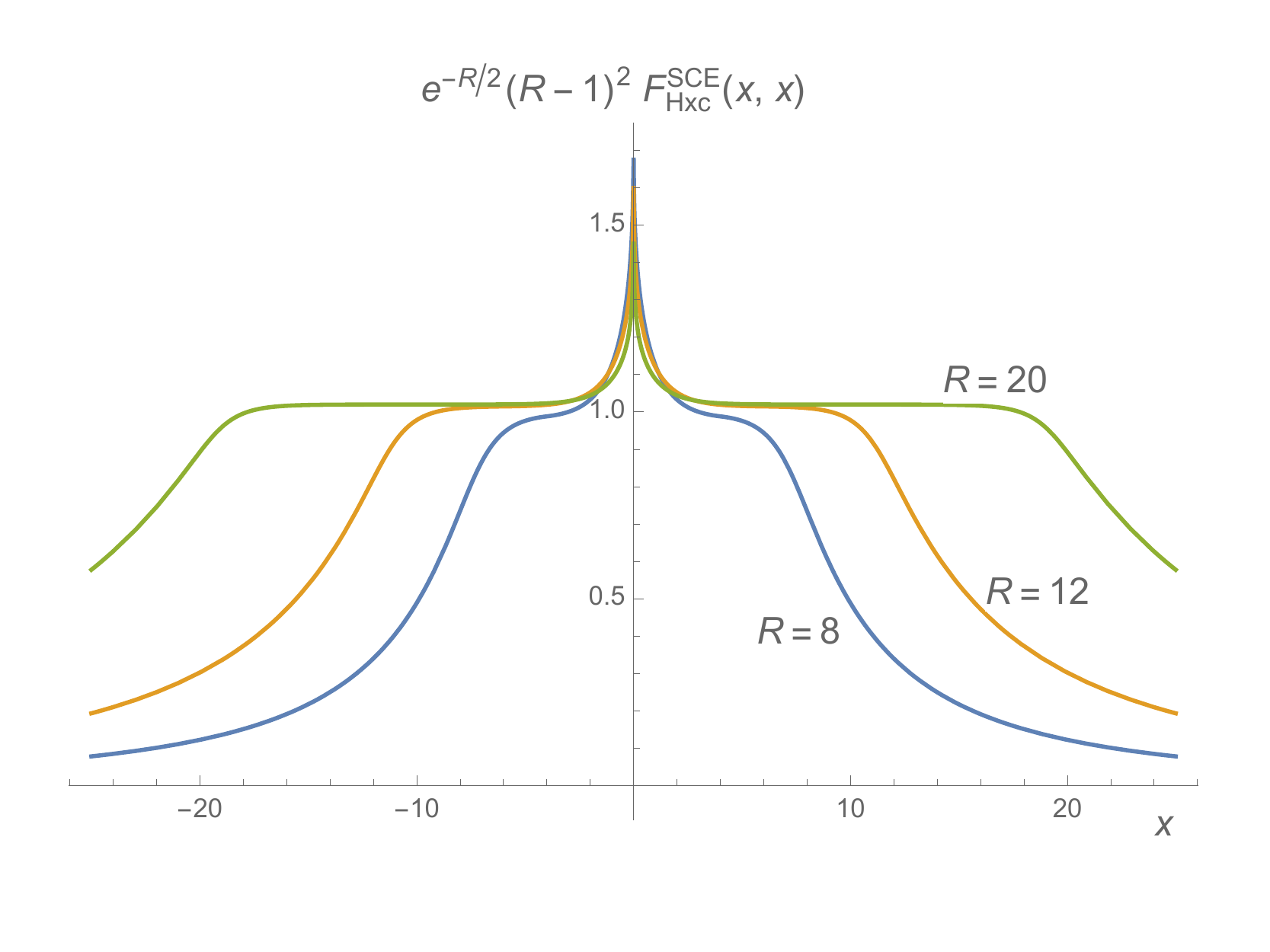}
 \caption{The SCE kernel for the ``homonuclear dimer'' density of Eq.~\eqref{eq:defdensH2} with $a=1$ and different internuclear separations $R$ along the diagonal $x=x'$. The kernel has been multiplied by $n(0)(R-1/a)^2$, to show its scaling with $R$. The value along the diagonal is exactly the same as the value of the kernel in the whole first and third quadrant, see Eq.~\eqref{eq:kernIandIII}  and Fig.~\ref{Fig:fig1}.}
 \label{Fig:plateau}
\end{figure}

The height of the plateaux increases as $R$ increases. A closer analysis of the function $G(x)$ defined by Eq.~\eqref{eq:defG} for the case of the density \eqref{eq:defdensH2} (see Appendix~\ref{App:G}) shows that the height and size of the plateaux are approximately given by
\begin{multline}
	\mathcal{F}^{\rm SCE}_{\rm Hxc}([n];x,x')\approx  \frac{1}{n(0)(R-1/a)^2} \\
	{\rm for}\;\, \frac{1}{a}\lesssim |x|\lesssim R-\frac{1}{a},\,\;{\rm and}\;\,\frac{1}{a}\lesssim |x'| \lesssim R-\frac{1}{a},\;\\
	{\rm with}\; 
	x,x'\ge 0\;\,{\rm or}\;\, x,x'\le 0
\label{eq:plateau}
\end{multline}
As an example, we show in Fig.~\ref{Fig:plateau} the SCE kernel along the diagonal for $R=8$, 12 and 20, multiplied by $n(0)(R-1/a)^2$, where $n(0)=e^{-a R/2}$. The value of the SCE kernel on the diagonal also defines the value of the kernel in the whole first and third quadrants, see Eq.~\eqref{eq:kernIandIII}.
Thus, we see that the SCE kernel develops plateaux regions whose height diverges exponentially as $R$ is increased.

This divergence is very promising to capture bond-breaking excitations, because it makes diverge matrix elements of the kernel between atomic orbitals centered {\it on the same site}.  Consider the basic example of the lowest excited singlet state $^1\Sigma_u^+$ of the H$_2$ molecule, \cite{GriGisGorBae-JCP-00,GieBae-CPL-08} where we have a matrix element of the kind
\begin{equation}
\int dx\int dx'\sigma_g(x)\sigma_u(x)\mathcal{F}^{\rm SCE}_{\rm Hxc}([n];x,x')\sigma_g(x')\sigma_u(x'),
	\label{eq:kernmatr}
\end{equation}
where $\sigma_{g,u}=c_{u,g} (\phi_A\pm \phi_B)$, with $\phi_{A,B}$ the atomic orbitals centered in the two atoms, and $c_{u,g}$ a normalization constant.
In the TDDFT linear response equations this matrix element is multiplied by the corresponding KS orbital energy difference $\epsilon_u-\epsilon_g$,
which goes to zero as $R\to\infty$, so that the kernel matrix element must diverge in order to keep the excitation finite (as it is in the exact system). \cite{GriGisGorBae-JCP-00,GieBae-CPL-08} We see that for the terms centered on the same atom $A$ appearing in \eqref{eq:kernmatr} we have, for large $R$,
\begin{multline}
	\label{eq:divmat}
	 \int dx\int dx'|\phi_A(x)|^2\mathcal{F}^{\rm SCE}_{\rm Hxc}([n];x,x')|\phi_A(x')|^2 \\ \approx \frac{1}{n(0)(R-1/a)^2}.
\end{multline}
Equation~\eqref{eq:divmat} holds because the product of the atomic orbitals with the same center, $|\phi_A(x)|^2|\phi_A(x')|^2$, is significantly different from zero only in one of the two plateau regions of Eq.~\eqref{eq:plateau}, where we can approximate the SCE kernel with the constant value $\frac{1}{n(0)(R-1/a)^2}$, which diverges exponentially as $R\to\infty$. Although a more careful analysis is needed to verify if this divergence is really able to compensate the vanishing of the KS excitation $\epsilon_u-\epsilon_g$ coming from a self-consistent KS SCE calculation, we see that the $\mathcal{F}^{\rm SCE}_{\rm Hxc}([n];x,x')$ embodies the right physics: its very non-local dependence on the density makes it diverge in the atomic region, {\em only when another distant atom is present}. 

Finally, the height of the peak in the origin can be easily obtained from the properties of the function $G(x)$ (see Appendix~\ref{App:G}), and it can be shown to be always equal to
\begin{equation}
	\mathcal{F}^{\rm SCE}_{\rm Hxc}([n];0,0)=2\mathcal{F}^{\rm SCE}_{\rm Hxc}\left([n];\frac{R}{2},\frac{R}{2}\right)\approx  \frac{2}{n(0)(R-1/a)^2}.
	\label{eq:peak}
\end{equation}
%

%
\section{Conclusions and perspectives}
In this work we have explored the SCE limit in the context of time-dependent problems, focusing on the formal properties of the adiabatic SCE (ASCE) functional.
We first examined some properties of the ASCE time-dependent potential, in particular the compliance to constraints of exact many-body theories, such as the generalized translational invariance and the zero-force theorem, and showed that the ASCE satisfies both.  While it is well known that non-locality in time requires non locality in space, we have shown that \textit{the converse is not true} using the example of the ASCE.

In the second half of the paper we derived an analytical expression for the SCE Hartree exchange-correlation kernel for one-dimensional problems, and we have computed it numerically for various density profiles. In particular, we have analyzed the case of a model homonuclear 2-electron molecule as the bond is stretched, finding that the SCE kernel displays a very promising diverging behavior that could tackle the problem of homolytic bond-breaking excitations.

In future works we will implement the whole linear response TDDFT equations for one-dimensional problems using the SCE kernel, analysing if its diverging behavior is able to open the gap in a model Mott insulator, made of a chain of H atoms. Work on bond-breaking excitations in real time propagation with the ASCE potential has also shown very promising results in this sense, and is currently in preparation. \cite{MirDegRubGor-XXX-16} Last but not least, we will use our insights to design approximate kernels based on the SCE form.
%
%
\section*{Acknowledgments}
It is our pleasure to dedicate this paper to Evert Jan Baerends, who has been a pioneer in understanding and using linear response TDDFT, and a wonderful mentor to some of us. We thank Andr\'e Mirtschink, Klaas Giesbertz and Michael Seidl for many useful discussions and for a critical reading of the manuscript.
This work was supported by an IEF fellowship from the Marie Curie program (n.~629625) and 
from the European Research Council under H2020/ERC Consolidator Grant ``corr-DFT'' (Grant No. 648932). RvL likes to thank the Academy of Finland for support. AG acknowledges CNPq for the financial support of his Ph.D., when this work was done.
%

%
\appendix
%
\section{Explicit calculation of the shifted co-motion functions in 1D}
 \label{App:1}
 Let us consider the negative semi-axis (the positive one gives an analogous result).
 The function $N_e([n];x)$ for a shifted density $n'$ reads:
 \begin{multline}
 N_{e}([n']; x) = \int\limits^{x}_{-\infty} n'(y) dy  = \int\limits^{x-R(t)}_{-\infty} n(y - R(t)) dy  \\
 = N_{e}([n]; x - R(t)), 
 \end{multline}
 and taking its inverse,
\bea
&& N_{e}^{-1} \left[N_{e}([n']; x)  \right] = x \nonumber \\
&&  N_{e}^{-1} \left[ N_{e}([n]; x - R(t)) \right] + R(t) = x. 
\eea
Combining the above relations 
we have:
\bea
\displaystyle
f_i([n'];x) &=& N_{e}^{-1} \left[ N_{e}([n']; x) + i-1 \right]  \\
&=& N_{e}^{-1} \left[ N_{e}([n]; x - R(t)) + i-1 \right] + R(t) \nonumber \\
&=& f_i([n]; x - R(t)) + R(t) \nonumber 
\eea
which is the 1D version of Eq.~(\ref{Eq:transf}).
\section{Functional variation of the 1D co-motion functions with respect to the density}
 \label{App:2}
Let $n$ be a density of a measure in $\R$ such that $\int_{\R} n(x) dx= N$. Then, the co-motion functions (or the optimal transport maps) are given explicitly by Eqs.~\eqref{Eq:como-1d}-\eqref{Eq:como-1d-2}, which correspond to the condition $ \int_x^{\ff_i(x)} n(y) dy \in \{ i, -N+i\}$, depending on whether $x \leq a_i$ or not. 

We consider $n^{\ep}(x) = n(x) + \ep (\tilde n(x) - n(x) )$ and we want to determine the corresponding co-motion functions $\ff^{\ep}_i$.  For every $x\in \R$ we define $x_{\ep}$ as the point such that $\ff_i (x_{\ep} ) = \ff_i^{\ep}(x)$, and then we notice that
\begin{multline}
\int_{x_{\ep}}^{\ff_i(x_{\ep})} n(y) dy - \int_{x}^{\ff_i^{\ep}(x)} n^{\ep}(y)dy  \in \{ -N , 0 , N \},\\
\int_{x_{\ep}}^{\ff_i^{\ep}(x)} n(y) dy - \int_{x}^{\ff_i^{\ep}(x)} n(y)- \ep \int_x^{\ff_i^{\ep}(x)} (\tilde{n}(y) -n(y) ) dy \\
 \in \{ -N , 0 , N \},\\
\int_{x_{\ep}}^{x} n(y) dy - \ep \int_x^{\ff_i^{\ep}(x)} (\tilde{n}(y) -n(y) ) dy  \in \{ -N , 0 , N \}.
\end{multline}
Since this last quantity is for sure less than $N$ in its absolute value, then it must be equal to $0$ and so
\begin{equation} 
	\label{eqn:approx_ep} 
	\int_{x_{\ep}}^x n(y) dy = \ep \int_x^{\ff_i^{\ep}(x)} (\tilde{n}(y) -n(y) ) dy;
\end{equation}
this implies for sure that if $n>0$ everywhere then $x_{\ep}-x$ is of order $\ep$ and then either $\ff_i - \ff_i^{\ep}$ is of order $\ep$ or (when $x=a_i$) $\ff_i \sim \pm \infty$ and $\ff_i^{\ep} \sim \mp \infty$. In the first case it is clear that the right-hand side of \eqref{eqn:approx_ep} can be approximated by $\int_x^{\ff_i(x)} (\tilde n - n)$ losing a lower order term; in fact this is true also in the second case, using the fact that $\tilde n - n$ has null total integral. Now, defining the quantity
\begin{equation}
 \alpha_i(x)= \int_x^{\ff_i(x)} \left(\tilde n(y) - n(y)\right) dy 
\end{equation}
and using \eqref{eqn:approx_ep}, the last observation, and the continuity of $n$, we can infer that
\begin{equation}
	 x_{\ep} = x - \ep \frac { \alpha_i(x) } {n(x)} + o(\ep).
\end{equation}
This means that we have
\begin{equation}
\frac{\ff_i^{\ep}(x)  - \ff_i (x) }{\ep} = \frac{\ff_i(x_\ep)  - \ff_i (x) }{\ep}  \sim - (\ff_i(x) ) ' \frac{ \alpha_i (x)}{n(x)}, 
\end{equation}
where the derivative is a distributional derivative, which takes into account also the jumps. Away from $x=a_i$, which is the only point where the jump can occur, we can make a simplification, as we have $\ff_i' (x) = n(x)/n(\ff_i(x))$, and so  $ (\ff_i(x) ) ' \frac{ \alpha_i (x)}{n(x)} = \frac{ \alpha_i (x)}{n(\ff_i(x))}$. For the jump we will get an additional term in the form of a delta function,
\begin{equation} 
 (\ff_i^+([n];a_i) - \ff_i^-([n];a_i) ) \delta (x- a_i) \cdot \frac{ \alpha_i (x)}{n(x)}.
\end{equation}
In the end, noticing that $-\alpha_i(x) =  \int_{\R} \delta n(x') \cdot \bigl( \theta(x-x') - \theta( \ff_i([n];x) - x') \bigr) dx'$, we get
\begin{multline}
 \frac { \delta \ff_i([n];x) }{\delta n (x')} =\bigl( \theta(x-x') - \theta( \ff_i([n];x) - x') \bigr) \cdot \\
\left( \frac{1 }{n (\ff_i([n];x)} + \frac{ \ff_i^+([n];a_i) - \ff_i^-([n],a_i )}{ n (a_i)} \cdot \delta(x- a_i)  \right)
\end{multline}

\section{Functional variation for the SCE potential (kernel)}
\label{App:varSCEpot}
Now we want to use the results of the previous section on the variation of the co-motion functions to derive an expression for $\delta v_{\rm Hxc}^{\rm SCE}(x) /\delta n(x')$, where $v_{\rm Hxc}^{\rm SCE}(x)$ is given in Eq.~\eqref{Eq:1dsce-pot}. We can write $\mathcal{F}_{\rm Hxc}^{\rm SCE}(x,x')=K_R(x,x')+K_S(x,x')$, where $K_R$ and $K_S$ are, respectively,  the contribution due to the regular and singular part of $\frac{\delta f_i(x)}{\delta n(x')}$. When the co-motion functions do not have jumps (that is, when $y \neq a_i$) we can simply apply the chain rule and find an expression for the regular part
\begin{multline}\label{eqn:first_term}  
K_R(x,x')=	\sum_{i=2}^N \int_x^{\infty} w''(|y-\ff_i([n];y)|)\cdot \\
 \frac{  \theta(y-x') - \theta( \ff_i([n];y) - x') }{n (\ff_i(y)) } dy.
\end{multline}
The singular part is more delicate as the classical chain rule does
not apply anymore: we find that $\delta \ff_i /\delta n$ is
proportional to $(\ff_i)'$. [The chain rule for derivatives when we
have a step function is not trivial: just take the example of $g (\theta)$, for which the derivative is $(g(1)-g(0)) \delta_0$ and not $(g' \circ \theta) \cdot \theta' = g'(0) \delta_0$.] In this case, the term coming from the jump, noticing that ${\rm sgn}(a_i-\ff^{\pm}_i([n];a_i)) = \mp 1 $, has the form
\begin{multline}
K_S(x,x')=\sum_{i=2}^N \theta( a_i-x) \cdot \Bigl(   w' (|a_i-\ff^+_i([n];a_i)|)  + \\
 w' (|a_i-\ff^-_i([n];a_i)|)  \Bigr) \cdot \frac {\alpha_i(a_i)}{n(a_i)}.
\end{multline}
In particular, whenever $n(x)>0$ on the whole real line we would have $\ff^{\pm} ([n]; a_i) = \pm \infty$ and this term becomes $0$ if $w'(\infty)=0$ which is the case for the Coulomb interaction or any situation in which the force tends to $0$ with the distance. For densities supported on the whole $\R$ we have only the term \eqref{eqn:first_term}, which coincides with Eq.~\eqref{Eq:1dker-general}.

However, if the support of $n$ is compact, denoting by $n^-, n^+$ the
extremes of the support, we would have $\ff^{\pm}([n];a_i)= n^{\pm}$
and the contribution would be nonzero. We can write it in a clearer form (using $\theta( \ff^+_i([n];a_i) - x') = 0$)
\begin{multline}\label{eqn:jump_term} 
	K_S(x,x')=\sum_{i=2}^N  \theta( a_i-x) \cdot \Bigl( w' (|a_i-n^+|) + w' (|a_i-n^-|)  \Bigr) \cdot\\
	 \frac { \theta(a_i-x') }{n(a_i)}.
\end{multline}

\section{Symmetry of the SCE kernel in 1D}
\label{app_symmetry}
The symmetry in $x$ and $x'$ of the singular term $K_S(x,x')$ of Eq.~\eqref{eqn:jump_term} is obvious, while for $K_R(x,x')$ of Eq.~\eqref{eqn:first_term} this is subtler. In order to prove it we compute $\frac{\partial^2}{\partial x \partial x'}K_R(x,x')$: if we can prove that this quantity is symmetric, then
the symmetry of the whole kernel is automatically proven: 
\begin{multline} 
\frac{\partial^2}{\partial x \partial x'} K_R ([n];x,x') =\\
 \sum_{i=2}^N w''(|x-\ff_i([n];x)|) \frac{ \delta(x' - x ) -\delta(x' - \ff_i([n];x))}{n(\ff_i([n];x))} \\
= \sum_{i=2}^N \frac{ w''(|x-\ff_i([n];x)|)  \delta(x' - x )}{n(\ff_i([n];x))}  -\\
 \sum_{i=2}^N \frac{ w''(|x-x'|) \delta(x' - \ff_i([n];x))
}{n(\ff_i([n];x))}.
\end{multline}
For the first term we will use the fact that $h(x) \delta(x'-x) = h(x') \delta(x-x')$, while for the second term we will use also that $\delta(g(x)-g(x'))=  \frac 1{g'(x)} \delta ( x- x')$. In particular we have $x'= \ff_i( \ff_{N-i+2} (x'))$, and so, since $\ff_i'([n],y) = \frac{ n(y)}{n(\ff_i([n];y)}$:
 \begin{multline}
	\delta(x' - \ff_i([n];x)) = \delta  ( \ff_i( \ff_{N-i+2} (x')) -  \ff_i(x) )  \\ 
	=\delta(\ff_{N-i+2}([n];x')-x) \frac {n(\ff_i([n];x))} { n(x) }.
\end{multline}
Plugging in this expression and using again the fact that $h(x) \delta(y-x) = h(y) \delta(x-y)$ we find that
\begin{multline} 
\frac{\partial^2}{\partial x \partial x'} K_R ([n];x,x') = \\
\sum_{i=2}^N \frac{ w''(|x-\ff_i([n];x)|)  \delta(x' - x )}{n(\ff_i([n];x))} \\
 - \sum_{i=2}^N \frac{ w''(|x-x'|) \delta(x' - \ff_i([n];x))}{n(\ff_i([n];x))} \\
= \sum_{i=2}^N \frac{ w''(|x'-\ff_i([n];x')|)  \delta(x - x' )}{n(\ff_i([n];x'))}\\
  - \sum_{i=2}^N \frac{ w''(|x-x'|) \delta(\ff_{N-i+2}([n];x') - x)
}{n(x)} \\
= \sum_{i=2}^N \frac{ w''(|x'-\ff_i([n];x')|)  \delta(x - x' )}{n(\ff_i([n];x'))} \\
  - \sum_{i=2}^N \frac{ w''(|x-x'|) \delta(\ff_{N-i+2}([n];x') - x)
}{n(\ff_{N-i+2}([n];x'))}  \\
= \frac{\partial^2}{\partial x \partial x'} K_R ([n];x',x),
\end{multline}
where in the last step we just relabeled the second sum.

\section{Properties of the function $G(x)$ for the homonuclear 1D density}
 \label{App:G}
 For the 2-electron density of Eq.~\eqref{eq:defdensH2} with $a=1$ it is easy to show that the co-motion function satisfies
\begin{equation}
	f([n];x\to 0^+)=\ln(x)-R+\ln\left(\frac{2}{1+e^{-R}}\right),
\end{equation}
yielding
\begin{equation}
	n(f([n];x\to 0^+))=x\, e^{-R/2}.
\end{equation}
The case $x\to 0^-$ can be obtained from $f([n];-x)=-f([n];x)$ and $n(-x)=n(x)$. Inserting these expansions in the definition of the function $G(x)$ of Eq.~\eqref{eq:defG} we see that its derivative is given by
\begin{equation}
	G'(x\to 0^+)=\frac{2\,e^{R/2}}{x|R-\ln\left(\frac{2}{1+e^{-R}}\right)-\ln(x)|^3},
\end{equation}
showing that $G(x)$ has an infinite slope in $x=0$. Furthermore, we also have, from the properties of the co-motion function and from the symmetry of the density $n(-x)=n(x)$ that
\begin{eqnarray}
	G(-x)& = & 2 \,G(0)-G(x) \label{eq:posG} \\
	G(f(x)) & = & G(x)- G(0) \qquad {\rm for}\;x>0\\
	G(f(x))& = & G(x)+G(0) \qquad {\rm for}\;x<0. \label{eq:negxG}
\end{eqnarray}
Since $f([n];-R/2)=R/2$, for $x=- R/2$ both properties \eqref{eq:posG} and \eqref{eq:negxG} must hold, implying that $G(0)=2G(-R/2)$, which is Eq.~\eqref{eq:peak}. \\
When $R$ is large, if $x$ is well inside one of the atomic regions then $f^{\mp}([n];x)\approx x\pm R$, yielding the constant distance $|x-f_i([n];x)|\approx R$, producing the plateaux regions in the kernel. This behavior holds until the electron in $f([n];x)$ approaches the origin and starts to ``see'' the second density in the overlap region present in the midbond. This happens when $x\approx \pm (R-1/a)$. At this point,  the large negative $x$ behavior of $G(x)$ starts to appear,
\begin{equation}
	G(x\to-\infty)=	\frac{1}{n(0) x^2},
\end{equation} 
yielding the plateau value of Eq.~\eqref{eq:plateau}.

%
%
\clearpage

 
%

\end{document}